# Light-Induced Giant Enhancement of the Nonlinear Hall Effect in Two-Dimensional Electron Gases at KTaO$_3$ (111) Interfaces

Hui Zhang,[1,2]* Daming Tian,[1] Xiaobing Chen,[3]* Weijian Qi,[1] Lu Chen,[1] Min Li,[1] Yetong Bai,[1] Jine Zhang,[1] Furong Han,[1] Huaiwen Yang,[1] Yuansha Chen,[4,5] Yunzhong Chen,[4,5] Jing Wu,[6,7] Yongbing Xu,[6,8] Fengxia Hu,[4,5,9] Baogen Shen,[4,5,9,10] Jirong Sun,[4,11]* and Weisheng Zhao[1,2]*

[1]School of Integrated Circuit Science and Engineering, Beihang University, Beijing 100191, China

[2]State Key Laboratory of Spintronics Hangzhou International Innovation Institute, Beihang University, Hangzhou 311115, China

[3]Quantum Science Center of Guangdong-Hong Kong-Macao Greater Bay Area (Guangdong), Shenzhen 518045, China

[4]Beijing National Laboratory for Condensed Matter Physics, Institute of Physics, Chinese Academy of Sciences, Beijing 100190, China

[5]School of Physical Sciences, University of Chinese Academy of Sciences, Beijing 100049, China

[6]York-Nanjing International Joint Center in Spintronics, School of Physics, Engineering and Technology, University of York, York YO10 5DD, UK

[7]School of Integrated Circuits, Guangdong University of Technology, Guangzhou 510006, China

[8]School of Integrated Circuits, Nanjing University, Suzhou 215163, China

[9]Songshan Lake Materials Laboratory, Dongguan, Guangdong 523808, China

[10]Ningbo Institute of Materials Technology & Engineering, Chinese Academy of Sciences, Ningbo, Zhejiang, 315201, China

[11]School of Physics, Zhejiang University, Hangzhou 310027, China

*Corresponding authors:
huizh@buaa.edu.cn
chenxiaobing@quantumsc.cn
jrsun@iphy.ac.cn
wszhao@buaa.edu.cn

**Abstract**

The nonlinear Hall effect (NLHE), an emergent phenomenon in noncentrosymmetric systems, enables the generation of a transverse voltage without an external magnetic field through a second-order electrical response. However, achieving a sizable NLHE signal remains a critical challenge for its application in frequency-doubling and rectifying devices. Here, we report a light-induced giant enhancement of the NLHE in the two-dimensional electron gas (2DEG) at the $CaZrO_3/KTaO_3$ (111) interface. Under light illumination, the second harmonic Hall voltage ($V_y^{2\omega}$) increases substantially and undergoes a sign reversal. Correspondingly, the second-order transverse conductivity ($\sigma_{yxx}^{(2)}$) increases by nearly five orders of magnitude, reaching 2.4 µm V$^{-1}$ Ω$^{-1}$, while also reversing its sign. Scaling analysis indicates that skew scattering is the dominant mechanism underlying the NLHE and is highly tunable via optical gating. Photoexcitation pumps electrons from in-gap states into the higher-lying Ta 5$d$ conduction band, generating high-mobility photocarriers that significantly increase the cubic transport scattering time ($\tau^3$), thereby driving a dramatic enhancement of $\sigma_{yxx}^{(2)}$. First-principles calculations further reveal that the Berry curvature distribution on the Fermi surface strongly depends on band filling. As the Fermi level approaches a band crossing in the Ta 5$d$ subband near the M point, the Berry curvature triple undergoes a sign change, accounting for the experimentally observed sign reversal of the nonlinear Hall response. Our work offers a new strategy to optically boost and tune the nonlinear Hall effect in oxide 2DEG systems, paving the way for applications in light-controlled rectification and nonlinear electronic devices.

## Introduction

The nonlinear Hall effect (NLHE), a recently discovered member of the Hall effect family, has attracted considerable interest as a second-order electrical response in time-reversal-invariant systems. Unlike the conventional Hall effect, when excited by an alternating current (AC) with a frequency $\omega$, the NLHE generates transverse Hall voltages at both zero (DC) and second-harmonic ($2\omega$) frequencies without requiring an external magnetic field[1-4]. This distinctive behavior opens new avenues for developing novel frequency-doubling and rectifying devices. Notably, the driving AC current can be replaced by an oscillating electromagnetic field, enabling a NLHE-based wireless rectifier with broadband response under zero bias, which holds great potential for applications in energy harvesting and wireless charging[5,6]. Recently, a pronounced NLHE has been observed in the two-dimensional electron gas (2DEG) at (111)-oriented amorphous-LaAlO$_3$/KTaO$_3$ (a-LAO/KTO) interface under zero magnetic field, where nonlinear transport is dominated by the skew scattering mechanism[7]. In addition to extrinsic origins such as inversion-asymmetric impurity potentials[8], skew scattering can also arise intrinsically from the chirality of Bloch wavefunctions, which is fundamentally linked to the Berry curvature distribution near the Fermi surface[7]. Owing to the broken inversion symmetry at the heterointerface, the strong spin-orbit coupling (SOC) arising from the heavy $5d$ element Ta[9-14], and the involvement of multiple $5d$ orbitals[15,16], KTO-based 2DEGs exhibit a hierarchical band structure with nontrivial Berry curvature textures, thereby providing an ideal platform for exploring nonlinear transport phenomena.

Due to strong quantum confinement at oxide interfaces, 2DEGs exhibit exceptional sensitivity to external fields, enabling efficient tunability of their electronic properties[15,17-19]. In particular, optical gating serves as a powerful means to modulate the carrier density (i.e., band filling) via photoexcitation, allowing precise control over the transport properties and emergent quantum phenomena of 2DEGs[10,20-22]. For example, a substantial enhancement of Rashba SOC under light illumination has been observed at a-Hf$_{0.5}$Zr$_{0.5}$O$_2$/KTO (110) heterointerfaces, where the Rashba effective field increases by a factor of seven[11]. Furthermore, a light-induced giant enhancement of nonreciprocal transport has been reported at (111)-oriented CaZrO$_3$/KTO (CZO/KTO) interfaces, with the nonreciprocal transport coefficient increasing by three orders of magnitude[16]. Given the strong Fermi-level dependence of the Berry curvature distribution at KTO-based interfaces[7], optical control of band filling provides an appealing route to modulate the nonlinear Hall response. However, optical tuning of the NLHE in 2DEGs at KTO interfaces remains largely unexplored.

In this work, we systematically investigated the NLHE of the 2DEG at the CZO/KTO (111) interface under illumination with light of varying power and wavelength. A light-induced giant enhancement of the NLHE is observed, manifested by a substantial increase of the second harmonic Hall voltage ($V_y^{2\omega}$) accompanied by a sign reversal. Notably, the corresponding second-order transverse conductivity ($\sigma_{yxx}^{(2)}$) exhibits a dramatic enhancement of nearly five orders of magnitude, increasing from $3.3\times10^{-5}$ μm V$^{-1}$ Ω$^{-1}$ in the dark to 2.4 μm V$^{-1}$ Ω$^{-1}$ under illumination, while also reversing its sign. This value far exceeds the previously reported optimal value among STO- and KTO-based 2DEGs ($|\sigma_{yxx}^{(2)}| = 7\times 10^{-2}$ μm V$^{-1}$ Ω$^{-1}$ in the a-LAO/KTO (001) system[7]), representing the highest $|\sigma_{yxx}^{(2)}|$ observed to date in oxide 2DEGs. It is also three orders of magnitude greater than the highest reported value for topological insulators ($|\sigma_{yxx}^{(2)}| = 2.4\times 10^{-3}$ μm V$^{-1}$ Ω$^{-1}$ in MnBi$_2$Te$_4$[23]). According to the scaling law, skew scattering is the dominant mechanism underlying the NLHE and can be effectively tuned through optical gating. Light illumination excites a second type of high-mobility photocarriers, which significantly increase the cubic transport scattering time ($\tau^3$), thereby driving a dramatic enhancement of $\sigma_{yxx}^{(2)}$. First-principles calculations reveal that the sign reversal of $\sigma_{yxx}^{(2)}$ upon Fermi level elevation is attributed to the sign change of the Berry curvature triple near a band crossing in the Ta 5$d$ subband. Our work not only reveals the critical role of photoexcitation in tuning nonlinear Hall transport via band filling and Berry curvature distribution engineering but also opens a new avenue for designing light-tunable rectification devices based on oxide 2DEGs.

**RESULTS AND DISCUSSION**

**1. Giant enhancement with sign reversal of the nonlinear Hall effect under irradiation**

The conductive interfaces were fabricated by growing CZO films on (111)-oriented KTO single-crystal substrates using the technique of pulsed laser deposition (see Methods section). A schematic illustration of the buckled honeycomb lattice structure of the KTO (111) surface is shown in Fig. 1a. Atomic force microscopy (AFM) imaging reveals an atomically flat surface morphology of the CZO film, with a root mean square roughness of ~2.2 Å (see Fig. S1). The bulk lattice parameter of CZO (pseudo-cubic $a_{CZO}$ = 4.012 Å) is very close to that of KTO ($a_{KTO}$ = 3.989 Å) with a lattice mismatch of only 0.58%. Such an excellent lattice match enables the epitaxial growth of the CZO/KTO (111) heterostructure, as demonstrated in previous work[16]. The transport properties were measured using a Hall-bar geometry (Fig. 1b), with detailed fabrication procedures described in the Methods section. A typical metallic behavior is observed across the temperature range of 2 to 300 K, confirming the formation of the 2DEG at the CZO/KTO (111) heterointerface, with a sheet carrier density of $n_S$ = 5.0×10$^{13}$ cm$^{-2}$ and a Hall mobility of $\mu$ = 212 cm$^2$ V$^{-1}$ s$^{-1}$ at 2 K (Fig. S2). To explore nonlinear transport under time-reversal symmetry, harmonic measurements were performed at zero magnetic field using low-frequency lock-in techniques, as schematically illustrated in Fig. 1b. A harmonic current ($I_\omega$) at a fixed frequency $\omega$ is applied along the Hall-bar channel, while the longitudinal ($V_x$) and transverse ($V_y$) voltages are simultaneously recorded at both the fundamental and second-harmonic frequencies. The influences of the diode effect, thermoelectric effect, and electrode misalignment on our measurements were carefully excluded (see Supplementary Note S1), confirming the absence of experimental artifacts. The photoresponse of the nonlinear transport was investigated by exposing the Hall-bar device to a laser source, as depicted in Fig. 1b. The heating effects induced by large current and laser illumination were ruled out (see Fig. S3), and the KTO substrate remained insulating during illumination, thereby ensuring the intrinsic nature of the 2DEG throughout the laser-irradiated transport measurements.

Figure 1c displays the nonlinear response measured with and without light illumination (wavelength $\lambda = 405$ nm and power $P = 0.45$ mW) for current applied along the [1-10] crystallographic direction at 10 K. As expected, sizable second harmonic Hall voltage $V_y^{2\omega}$ outputs are clearly observed in the absence of an applied magnetic field. All measured $V_y^{2\omega}$, under both dark and illuminated conditions, exhibit a well-defined quadratic dependence on the applied current $I_\omega$, and reverse sign when simultaneously reversing the current direction and voltage probe electrodes, as schematically shown in the inset of Fig. 1c, where $I_\omega$ is applied in the positive (red) or negative (black) directions. This feature is indicative of a hallmark of nonlinear transport. In the dark, $V_y^{2\omega}$ is measured to be +38 μV at $I_\omega = +100$ μA, with its magnitude and sign being consistent with the result previously reported at the a-LAO/KTO (111) interface[7]. Strikingly, under irradiation, $V_y^{2\omega}$ undergoes a sign reversal compared to the dark condition under the same measurement configuration and exhibits a dramatic increase in magnitude by a factor of 11, reaching −424 μV at $I_\omega = +100$ μA. This result highlights the remarkable tunability of the nonlinear Hall response by optical stimuli. We also measured the nonlinear Hall voltage $V_y^{2\omega}$ with the current applied along [11-2] crystallographic directions of the KTO (111) substrate (Fig. S4). In contrast, a much smaller $V_y^{2\omega}$ was detected in the dark at 10 K. As reported, a similar anisotropy in the NLHE response has also been observed at the a-LAO/KTO (111) interface[7]. Furthermore, pronounced enhancement with sign reversal of the NLHE is also observed under light illumination when the current is applied along the [11-2] orientation. In the following discussion, unless otherwise noted, $I_\omega$ is applied in the positive direction along the [1-10] crystallographic orientation. Under an AC excitation current $I_\omega = I_0 \sin(\omega t)$, the voltage response follows $V_y \propto [I_0\sin(\omega t)]^2 = I_0^2[1+\sin(2\omega t-\pi/2)]/2$, indicating that the coexistence of a DC and a second-harmonic component is a generic feature of the second-order transport effects[2]. Accordingly, DC measurements were carried out, and a DC Hall voltage $V_y^{DC}$ that scales quadratically with the applied AC current was clearly observed both with and without light illumination, exhibiting an amplitude comparable to that of the second-harmonic component (see Fig. S5). Additionally, to further validate our experimental findings, we conducted Hall measurements with input currents at different frequencies and found that the second-harmonic signal is independent of the AC frequency under both dark and illuminated conditions, as shown in Fig. 1d. All these observations unambiguously demonstrate the existence of a transverse nonlinear electrical response under time-reversal symmetry at the CZO/KTO (111) interface, where the NLHE exhibits a giant enhancement accompanied by a sign reversal under irradiation, highlighting the significant role of optical control.

The temperature dependence of the light-induced modulation of NLHE performance was also investigated, and the $V_y^{2\omega}$ results with and without light irradiation at various temperatures are shown in Fig. 1e. For clarity, only $V_y^{2\omega}$ data from 10 K to 100 K are presented, while the complete $V_y^{2\omega}$ data up to room temperature and the corresponding first-harmonic longitudinal voltage $V_x^\omega$ are provided in Fig. S7. It is evident that $V_y^{2\omega}$ scales quadratically with $I_\omega$ across all the measured temperatures. In the dark, the output $V_y^{2\omega}$ value is positive and increases with rising temperature, consistent with the behavior observed at the a-LAO/KTO (111) interface[7]. Under irradiation, as the temperature increases from 10 K to 50 K, the sign of $V_y^{2\omega}$ is negative and its magnitude first increases and then decreases, but it is still lager than that in the dark condition. However, when the temperature exceeds 50 K, the sign of $V_y^{2\omega}$ reverses, and its magnitude becomes almost identical to that in the dark condition. These results indicate that the effect of photoexcitation on the NLHE is prominent below 50 K, consistent with the light-induced drop in sheet resistance ($R_S$) observed at low temperatures (Fig. S8). The suppressed photoresponse of the NLHE at higher temperatures may be attributed to enhanced thermal fluctuations.

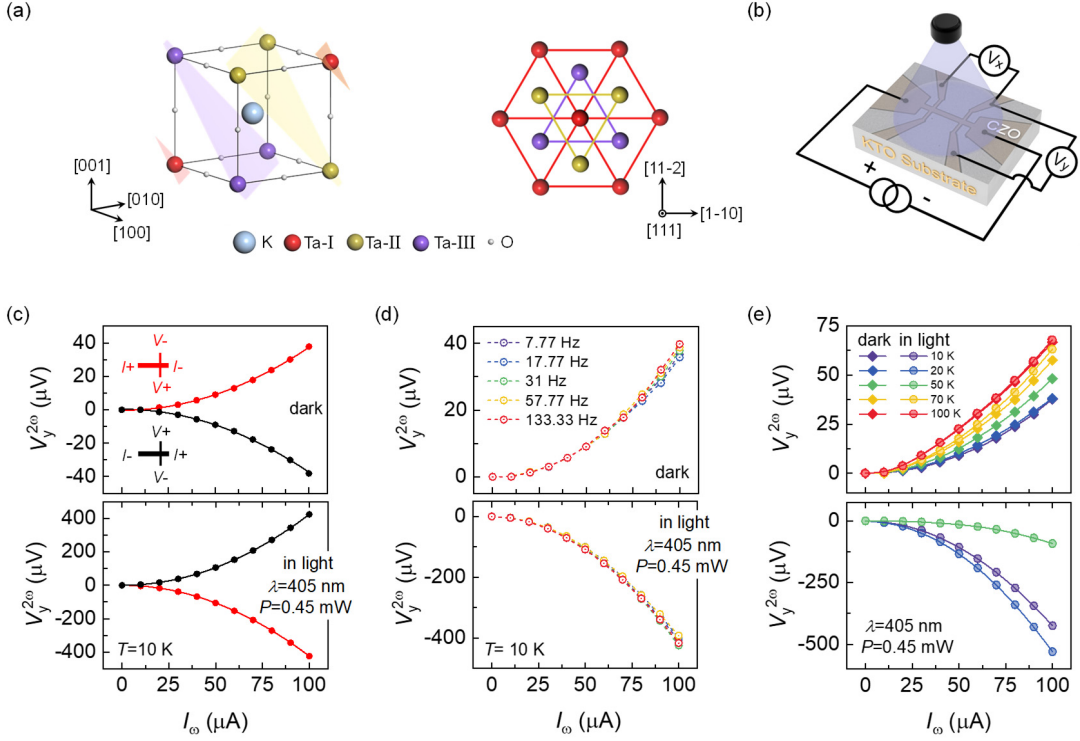

**Fig. 1 | Light-induced giant enhancement and sign reversal of the nonlinear Hall response.** (a) Schematic illustration of the perovskite KTaO$_3$ unit cell, with three adjacent (111) planes containing Ta$^{5+}$ ions shaded in light red, yellow, and purple, respectively (left panel). The right panel shows a top view of the Ta$^{5+}$ ion distribution along the [111] crystal axis with Ta-I, Ta-II, and Ta-III labeled accordingly. (b) Photograph of a typical Hall bar device for the CZO/KTO (111) heterostructure and a schematic of the nonlinear Hall measurement configuration under illumination. (c) The second harmonic transverse voltage $V_y^{2\omega}$ as a function of the AC current amplitude $I_\omega$ at 10 K, measured without (top panel) and with (bottom panel) light illumination (wavelength $\lambda = 405$ nm and power $P = 0.45$ mW). The insets illustrate the measurement geometry, where both the current direction and voltage probe electrodes are simultaneously reversed, with $I_\omega$ applied in the positive (red) or negative (black) directions. (d) $V_y^{2\omega}$ versus $I_\omega$ curves measured at various $I_\omega$ frequencies under the positive current direction. (e) $V_y^{2\omega}$ as a function of $I_\omega$ with and without light illumination at different temperatures. The solid lines in (c) and (e) are quadratic fits to the experimental data.

**2. Laser-tunable nonlinear Hall effect and high-mobility photocarrier generation**

To gain a comprehensive understanding of the photoresponse of the 2DEG at the CZO/KTO (111) heterointerface, systematic optical gating experiments were conducted under laser excitation with wavelength $\lambda$ ranging from 375 to 660 nm. Since the corresponding photon energy (1.88~3.32 eV) is lower than the band gap of KTO (~3.58 eV), light illumination can excite electrons from in-gap states, originating from oxygen vacancies, into the conduction band minimum of KTO, thereby inducing charge carrier pumping (Fig. 2a)[10,22,24]. As depicted in Fig. 2b and 2c, the first-harmonic longitudinal voltage $V_x^\omega$ and second harmonic transverse voltage $V_y^{2\omega}$ were measured as functions of the applied AC current $I_\omega$ at 10 K under irradiation ($\lambda$ = 405 nm) with laser power $P$ varying from 0 to 15 mW. The linear dependence of $V_x^\omega$ on $I_\omega$ indicates the formation of good ohmic contacts, and its monotonically decreasing slope with increasing $P$ reflects a pronounced photoinduced reduction in the sheet resistance $R_S$ of the 2DEG. This is corroborated by the dependence of $R_S$ on $P$ measured at 10 K, which shows that $R_S$ decreases by ~80%, from 514 $\Omega/\square$ in the dark to 99 $\Omega/\square$ at $P$ = 15 mW (Fig. S9). In Fig. 2c, the quadratic dependence of $V_y^{2\omega}$ on $I_\omega$ is observed across all $P$ values. The evolution of $V_y^{2\omega}$ with increasing $P$ shows no significant change within the range of 0 to 0.05 mW, with only a slight increase in magnitude due to the low laser intensity. However, once $P$ reaches the critical threshold of 0.1 W, the sign of $V_y^{2\omega}$ reverses, and its magnitude increases steadily as $P$ further grows. To better illustrate the tuning effect, we plotted the extracted $V_y^{2\omega}$ at a fixed input current $I_\omega$ = +60 µA as a function of $P$ (see the inset of Fig. 2c). In the dark ($P$ = 0), $V_y^{2\omega}$ = +13 µV, while at $P$ = 15 mW, $V_y^{2\omega}$ = −840 µV, corresponding to a modulation by a factor of over 65 in magnitude, highlighting the superior capability of optical gating in manipulating the nonlinear Hall response.

To further explore the laser wavelength dependence of the nonlinear response, we fixed the laser power at $P$ = 6 mW and varied the laser wavelength $\lambda$ from 660 to 375 nm. The first-harmonic longitudinal voltage $V_x^\omega$ decreases with decreasing $\lambda$, in agreement with the $\lambda$-dependent modulation of the sheet resistance $R_S$ (Fig. S10). A shorter $\lambda$ (i.e., higher photon energies) is expected to enhance charge carrier excitation, leading to a more pronounced reduction in $R_S$. Figure 2d presents the quadratic dependence of $V_y^{2\omega}$ on $I_\omega$, measured at 10 K in the dark and under illumination at various wavelengths with a fixed power of 6 mW. Compared to the dark condition, $V_y^{2\omega}$ exhibits a substantial increase under illumination at $\lambda$ = 660 nm and 532 nm, indicating an enhanced nonlinear response driven by photoexcitation. Specifically, $V_y^{2\omega}$ increases from +13 µV in the dark to +161 µV at $\lambda$ = 660 nm and +224 µV at $\lambda$ = 532 nm for $I_\omega$ = +60 µA. However, as the $\lambda$ further reduces to 405 nm and 375 nm, $V_y^{2\omega}$ not only reverses sign but also exhibits a significantly larger magnitude, reaching $V_y^{2\omega}$ = −664 µV at $\lambda$ = 405 nm and $V_y^{2\omega}$ = −517 µV at $\lambda$ = 375 nm for $I_\omega$ = +60 µA. Thus, as $\lambda$ decreases from 660 nm to 375 nm, $V_y^{2\omega}$ undergoes a sign reversal and a pronounced magnitude enhancement. Notably, whether the laser power or wavelength is varied or not, $V_y^{2\omega}$ reverses sign when the current direction and the corresponding Hall probes are inverted, and $V_y^{2\omega}$ remains independent of the AC frequency (see Fig. S11). These results confirm the reliability of the second-order nonlinear electrical response under laser irradiation across varying power levels and wavelengths. Therefore, our results clearly demonstrate that optical gating, achieved by tuning either the laser power or the wavelength, serves as an effective approach for controlling the NLHE of the 2DEG at the CZO/KTO (111) interface.

Next, we systematically investigate the effect of photoexcitation on the transport properties. The Hall resistance $R_{xy}$ as a function of the magnetic field $B$ at 10 K under 405 nm laser illumination with varying $P$ is shown in Fig. 2e. A well-defined linear $R_{xy}$-$B$ relation is observed for $P$ ranging from 0 to 0.05 mW (inset of Fig. 2e), which is a typical feature of 2DEGs with a single type of charge carriers. The slight decrease in the slope of the $R_{xy}$-$B$ curve with increasing $P$ suggests photoexcitation of electrons from in-gap states, resulting in an increased carrier density. However, when $P \geq 0.1$ mW, the Hall effect undergoes a linear to nonlinear transition, developing further with the increase of light intensity. Such a light-induced nonlinear Hall effect has also been observed in previously reported KTO-based 2DEGs, which is attributed to the photoexcitation of an additional high-mobility electron channel[10,11]. To gain insights into the electron excitation process, a two-band model is employed[25], as detailed in the Supplementary Note S2. As evidenced by the black solid fitting curves in Fig. 2e, the two-band model provides a satisfactory description to the nonlinear Hall effect. The extract carrier density and the corresponding Hall mobility are plotted as functions of laser power $P$ in Figs. 2f and 2g, respectively. In the region $P<0.1$ mW, the intrinsic electrons ($n_{S1}$) exclusively dominate the conductivity with a Hall mobility $\mu_1$ of approximately 200 cm$^2$ V$^{-1}$ s$^{-1}$. Notably, the light-induced second type of electrons ($n_{S2}$) appears when $P \geq 0.1$ mW, exhibiting a high Hall mobility ($\mu_2$). As $P$ increases from 0.1 to 15 mW, $n_{S2}$ rises from 0.63×10$^{12}$ to 5.26×10$^{12}$ cm$^{-2}$, while the corresponding Hall mobility $\mu_2$ increases from 2403 to 9060 cm$^2$ V$^{-1}$ s$^{-1}$, representing a 43-fold enhancement relative to the dark condition (~211 cm$^2$ V$^{-1}$ s$^{-1}$). It is worth noting that the light-induced enhancement with sign reversal of $V_y^{2\omega}$ for $P \geq 0.1$ mW (Fig. 1c) coincides with the emergence of the nonlinear Hall effect, suggesting a strong correlation with the emergence of high-mobility photoexcited carriers. The temperature dependence of Hall effect under 405 nm laser illumination at $P$ = 0.45 nm further confirms that such high-mobility carriers emerge below 50 K (Fig. S12), and in this temperature range, photoexcitation also has a pronounced impact on the NLHE (Fig. 1e). In addition, the Hall effect under illumination with various laser wavelengths was measured at $T$ = 10 K (Fig. S12). The dependence of carrier density and Hall mobility on $\lambda$ at a fixed power of 6 mW is presented in Figs. 2h and 2i, respectively. At wavelengths of 660 nm and 535 nm, only one type of carrier exists, corresponding to the positive $V_y^{2\omega}$ observed in Fig. 1d, whereas at wavelengths of 405 nm and 375 nm, the presence of high-mobility charge carriers corresponds to the negative $V_y^{2\omega}$.

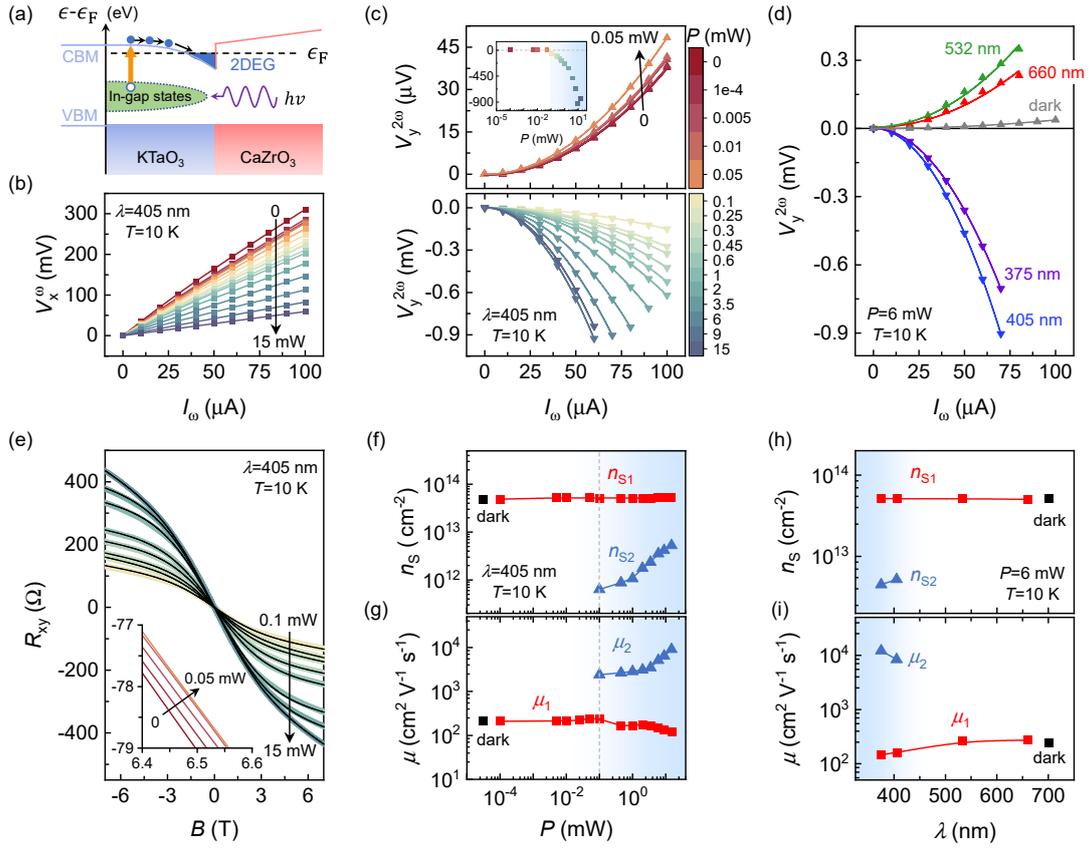

**Fig. 2 | Tuning of NLHE and transport behavior via optical gating.** (a) The schematic of the optical gating mechanism. (b) First-harmonic longitudinal voltage $V_x^\omega$ and (c) second harmonic transverse voltage $V_y^{2\omega}$ as functions of applied AC current $I_\omega$, measured at 10 K under light illumination ($\lambda$ = 405 nm) with laser power $P$ varying from 0 to 15 mW. The solid lines in (b) and (c) represent linear and quadratic fits to the data, respectively. (d) $V_y^{2\omega}$ versus $I_\omega$ in the dark and under illumination at different laser wavelengths at a fixed power of $P$ = 6 mW measured at 10 K. The solid lines are quadratic fits to the data. (e) The evolution of Hall effect with increasing laser power at 10 K under laser illumination ($\lambda$ = 405 nm). The solid black curves are the best fits to the two-band model. (f, g) Extracted carrier density and Hall mobility, respectively, as functions of laser power, obtained from the two-band model employed to the experimental data in (e). (h, i) Carrier density and Hall mobility, respectively, as functions of laser wavelength at a fixed power $P$ = 6 mW measured at 10 K. Solid lines in (f)-(i) are guides for the eye.

## 3. Origin of the giant enhancement with sign reversal of the NLHE

The well-known mechanism underlying the NLHE is the Berry curvature dipole (BCD), which originates from inversion symmetry breaking that segregates positive and negative Berry curvatures in different momentum regions, resulting in a dipole moment when the crystalline symmetry allows[1]. However, the threefold rotational symmetry of the KTO (111) surface (Fig. 1a) prohibits the formation of a BCD[8,26]. As a result, the BCD-induced NLHE is forbidden in the 2DEG at the CZO/KTO (111) interface. Note that transverse nonlinear transport can arise not only from the intrinsic BCD but also from extrinsic effects, such as skew scattering and side jump[27-29]. Among them, skew scattering stems from the asymmetry of the differential scattering cross section due to the inherent chirality of the Bloch electrons, which is induced by the inversion symmetry breaking[8,30,31]. As depicted in Fig. 3a, the wave packet of the Bloch electron self-rotates due to a finite Berry curvature, and its chirality reverses with a change in the sign of the Berry curvature. When the self-rotating wave-packet is scattered by an impurity center, the asymmetry of the scattering cross section can give rise to a rectified nonlinear Hall voltage[8]. Thus, the skew scattering related nonlinear transport can serve as a hallmark of Berry curvature in KTO-based 2DEGs.

To explore the physical origin of the observed light-induced giant enhancement with sign reversal of nonlinear transport at the CZO/KTO (111) interface, the second-order transverse conductivity $\sigma_{yxx}^{(2)}$ is introduced to characterize the strength of the nonlinear response, independent of sample geometry. The linear longitudinal and nonlinear transverse conductivities $\sigma_{xx}$ and $\sigma_{yxx}^{(2)}$ are related through the second-order current density expression: $J_y^{(2)} = \sigma_{yxx}^{(2)}(E_x)^2 = \sigma_{xx}E_y^{(2)}$, where $E_x = V_x^\omega / L$ and $E_y^{(2)} = V_y^{2\omega}/W$ represent the first harmonic longitudinal and second harmonic transverse electric fields, respectively. $L$ and $W$ denote the length and width of the Hall-bar device. Accordingly, the nonlinear transverse conductivity $\sigma_{yxx}^{(2)}$ can be extracted from the above measurements using the formula: $\sigma_{yxx}^{(2)} = (\sigma_{xx}V_y^{2\omega}L^2)/[(V_x^\omega)^2 W]$. Based on this expression, we further analyze the scaling behavior of the second-order transport with respect to $\sigma_{xx}$ using the experimental data from Figs. 1e, 2d, and 2e (The variation of $\sigma_{yxx}^{(2)}$ with laser power, temperature, and wavelength is shown individually in Fig. S13). In Fig. 3b, $\sigma_{yxx}^{(2)}/\sigma_{xx}$ is plotted as a function of the squared longitudinal conductivity $\sigma_{xx}^2$, with data points classified into positive and negative groups according to the sign of $\sigma_{yxx}^{(2)}$. In both cases, $\sigma_{yxx}^{(2)}/\sigma_{xx}$ scales linearly with $\sigma_{xx}^2$, and the experimental data fit well with the relation[8]:

$$\sigma_{yxx}^{(2)}/\sigma_{xx} = \xi\sigma_{xx}^2 + \eta \quad (1)$$

where the coefficients $\xi$ and $\eta$ correspond to contributions to $\sigma_{yxx}^{(2)}$ that scale as $\sigma_{xx}^3$ and $\sigma_{xx}$, respectively. A magnified view of the positive data group is shown in the inset of Fig. 3b. Given that $\sigma_{xx}$ is linearly dependent on the transport scattering time $\tau$, thereby the $\sigma_{yxx}^{(2)}$ contains two components scaling as $\tau^3$ and $\tau$. The slope $\xi$ quantifies the $\tau^3$-dependence contribution originating from skew scattering, while the intercept $\eta$ accounts for the $\tau$-linear contribution associated with mechanisms including the BCD and side jump[8]. In our system, however, the BCD is not allowed due to the threefold rotational symmetry. Obviously, the $\tau$-linear term is attributed to the side jump mechanism. Therefore, the two contributors to the nonlinear response in 2DEG at the CZO/KTO (111) interface are skew scattering and side jump.

In the low-$\sigma_{xx}^2$ regime ($\sigma_{xx}^2 \leq 5\times10^{-6}\ \Omega^{-2}$), where only a single type of charge carrier is present, the values of $\sigma_{yxx}^{(2)}/\sigma_{xx}$ are positive. A linear fit to the data yields a slope of $\xi = +3.51\times10^4\ \mu m\ V^{-1}\ \Omega^2$ and an intercept of $\eta = +0.03\ \mu m\ V^{-1}$. Notably, as $\sigma_{xx}^2$ increases beyond this regime, $\sigma_{yxx}^{(2)}/\sigma_{xx}$ becomes negative, corresponding to a reversal in the sign of $V_y^{2\omega}$ when a second type of high-mobility charge carrier is excited by light illumination. This transition is accompanied by a sign change in the slope $\xi$, which becomes negative with a value of $-2.49\times10^6$ μm V$^{-1}$ Ω$^2$, nearly two orders of magnitude larger in absolute value than that in the low-$\sigma_{xx}^2$ regime. Meanwhile, the intercept $\eta$ remains positive but increases substantially to $+13.9$ μm V$^{-1}$. Using these fitting parameters, the distinct contributions to $\sigma_{yxx}^{(2)}$ can be extracted and plotted as a function of $\sigma_{xx}$ in Fig. 3c. The solid curves correspond to the cubic components attributed to skew scattering, while the dashed lines represent the linear side jump contributions, clearly indicating that the skew scattering mechanism dominates the transverse nonlinear response. More importantly, both the magnitude and sign of the skew scattering contribution are highly tunable via optical excitation, highlighting the crucial role of photo-induced carriers in modulating the NLHE.

Figure 3d displays the absolute value of the second-order transverse conductivity $|\sigma_{yxx}^{(2)}|$ as a function of $\sigma_{xx}$. As $\sigma_{xx}$ increases from $7.3\times10^{-4}$ to $1\times10^{-2}$ Ω$^{-1}$, $|\sigma_{yxx}^{(2)}|$ rises monotonically from $3.3\times10^{-5}$ to $2.4$ μm V$^{-1}$ Ω$^{-1}$, indicating an enhancement of nearly five orders of magnitude. The second-order transverse conductivity arising from skew scattering is given by

$$\sigma_{yxx}^{(2)} = \frac{e^3 v \tau^3}{\hbar^2 \tilde{\tau}} \tag{2}$$

where $\tau$ is the transport scattering time, $\tilde{\tau}$ is the skew scattering time, $v$ is the Dirac velocity, $e$ is the electric charge, and $\hbar$ is the reduced Planck constant. As indicated by this expression, $\tau$ plays a critical role in determining the magnitude of $|\sigma_{yxx}^{(2)}|$ due to its cubic dependence. Since $\tau \propto \mu$, the cube of the total Hall mobility $\mu_{total}^3$ is also plotted in Fig. 3d as a function of $\sigma_{xx}$. Here, $\mu_{total}^3$ is determined from the experimental data in Figs. 2g, 2i, and S2 via $\mu_{total} = \frac{\sigma_{xx}}{e n_{total}}$, where $n_{total} = n_{s1}$ when there is only one species of sheet carriers and $n_{total} = n_{s1}+n_{s2}$ when two species of charge carriers coexist. Notably, $\mu_{total}^3$ increases monotonically from $1.8\times10^5$ to $8.3\times10^8$ cm$^6$ V$^{-3}$ s$^{-3}$ over the same $\sigma_{xx}$ range. This close correspondence strongly supports the conclusion that the pronounced increase in $\mu_{total}^3$ (i.e., $\tau^3$) due to the photoexcitation of high-mobility carriers under illumination contributes to the substantial enhancement of $|\sigma_{yxx}^{(2)}|$.

In Fig. 3e, we summarize the dependence of the absolute values of $\sigma_{yxx}^{(2)}$ on $\sigma_{xx}$ for our 2DEG at the CZO/KTO(111) interface under optical gating and other typical NLHE systems previously reported, including oxide 2DEGs[7,32] topological insulators (TIs)[8,23] and 2D/Weyl materials[2,4,6,33-35]. The maximum $|\sigma_{yxx}^{(2)}|$ observed in our work exceeds that of most materials and is even three orders of magnitude greater than the highest reported value for topological insulators ($|\sigma_{yxx}^{(2)}| = 2.4\times10^{-3}$ μm V$^{-1}$ Ω$^{-1}$ in MnBi$_2$Te$_4$ (6-SL)[23], where SL denotes septuple layers). Although only surpassed by the hBN/graphene/hBN moiré superlattice, the maximum $|\sigma_{yxx}^{(2)}| = 2.4$ μm V$^{-1}$ Ω$^{-1}$ achieved in our work far exceeds the previously reported optimal value among STO- and KTO-based 2DEGs ($|\sigma_{yxx}^{(2)}| = 7\times10^{-2}$ μm V$^{-1}$ Ω$^{-1}$ in the a-LAO/KTO (001) system[7], representing the highest $|\sigma_{yxx}^{(2)}|$ observed to date in oxide 2DEGs. Notably, for the 2DEG at the a-LAO/KTO (111) interface, $|\sigma_{yxx}^{(2)}|$ can be tuned by nearly one order of magnitude via electrostatic gating. In contrast, under our optical gating, $|\sigma_{yxx}^{(2)}|$ at the CZO/KTO (111) interface exhibits an extensive range of tunability, spanning nearly five orders of magnitude. This comparison highlights the exceptionally large and optically tunable nonlinear Hall response demonstrated in our work, shedding light on the promising potential of KTO-based interfaces for applications in light-controlled wireless radiofrequency rectification devices.

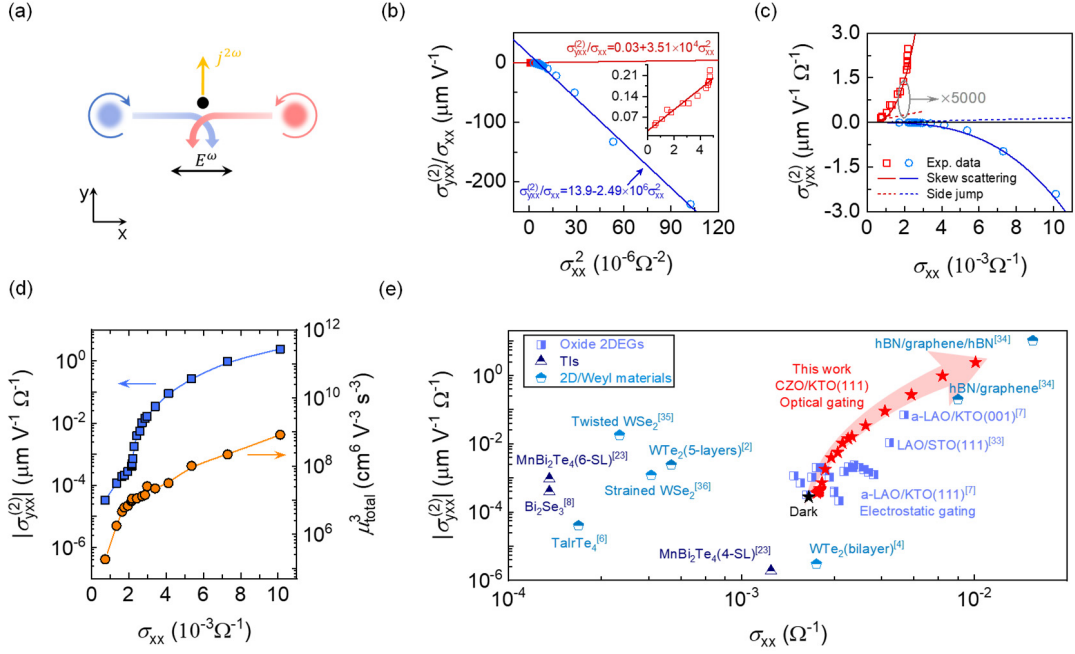

**Fig. 3 | Physical origin and scaling law of second-order transverse conductivity.** (a) Schematic illustration of skew scattering leading to nonlinear transverse transport. Blue and red spheres represent self-rotating electron wave packets with opposite chirality (Berry curvature). (b) $\sigma_{yxx}^{(2)}/\sigma_{xx}$ plotted as a function of the squared longitudinal conductivity $\sigma_{xx}^2$, using experimental data from Figs. 1e, 2d, and 2e. Data points are classified into positive and negative groups according to the sign of $\sigma_{yxx}^{(2)}$, and the solid lines represent linear fits to the experimental data. The inset shows a magnified view of the positive data group. (c) Decomposition of $\sigma_{yxx}^{(2)}$ into distinct contributions as a function of $\sigma_{xx}$. The dashed lines represent the linear side jump contribution, while the solid curves correspond to the cubic skew scattering components. (d) Absolute values of the nonlinear transverse conductivity $|\sigma_{yxx}^{(2)}|$ and $\mu_{total}^3$ as a function of $\sigma_{xx}$. Solid lines are guides for the eye. (e) Summary of the absolute values of $\sigma_{yxx}^{(2)}$ for our 2DEG at the CZO/KTO (111) interface and other typical systems previously reported, including oxide 2DEGs[7,32], topological insulators (TIs)[8,23] and 2D/Weyl materials[2,4,6,33-35], where SL denotes septuple layers. This comparison highlights the exceptionally large and optically tunable nonlinear Hall response in our system.

Photoexcitation can alter the occupation of electronic states by shifting the Fermi level upward. To further elucidate the effect of band filling on nonlinear transport, density functional theory (DFT) calculations were performed using the model described in the Methods section. The electronic states near the Fermi level in KTO originate from the tantalum $5d$ orbitals. When tantalum is surrounded by an oxygen octahedron, its $5d$ orbitals are split by the cubic crystal field into a higher-energy doublet ($e_g$ states) and a lower-energy triplet ($t_{2g}$ states). The geometry of the (111) KTO surface further breaks the $t_{2g}$ orbitals into an $a_1$ singlet and a two-fold degenerate $e'$ ($e'_1$ and $e'_2$) doublet due to the trigonal crystal field under $C_{3v}$ symmetry. When SOC is considered, the $e'$ orbitals are further coupled into states with angular momentum projections $L_{z,-}$ and $L_{z,+}$. The resulting energy-level hierarchy, supported by DFT calculations, is schematically illustrated in Fig. 4a (see Supplementary Note S3 for details). Figure 4b presents the calculated electronic band structure of the 2DEG residing on the KTO (111) surface, where pronounced Rashba spin splitting is observed near the $\Gamma$ point. For comparison, the band structure with and without SOC are shown in Supplementary Fig. S14. In contrast to the canonical Rashba interaction which exhibits linear dependence on wavevector $k$, the doublet splitting shows a trigonal warping with cubic dependence.

Representative Fermi surfaces and the corresponding Berry curvature distributions at different Fermi levels $\epsilon_F$ = 0.06, 0.24 and 0.27 eV are shown in Fig. 4c, with the Fermi levels marked by the black, blue and red dashed lines in Fig. 4b. Under dark condition, the electronic band filling is relatively low, and only one type of electron occupies the lower $L_{z,-}$ orbital. Upon illumination, photoexcitation increases the carrier density, causing the Fermi level to shift upward and cross the bottom of the upper $L_{z,+}$ orbital. The population of photocarriers in the $L_{z,+}$ bands leads to the emergence of a second species of charge carriers. As discussed above, these photoexcited electrons exhibit remarkably high mobility (Fig. 2), leading to a significantly enhanced cubic transport scattering time ($\tau^3$), and consequently a giant increase in the magnitude of $\sigma^{(2)}_{yxx}$. While the threefold symmetry suppresses the Berry curvature dipole (BCD), it provides a trigonal warped Fermi surface essential for the Berry curvature triple (BCT), a higher-order moment of the Berry curvature distribution in momentum space that enables skew scattering to dominate the nonlinear transport[8,30,31]. The Berry curvature triple $T(\epsilon_F)$ quantifies the strength of the Berry curvature $\Omega_z(k)$ on the Fermi surface, and can be described as: $T(\epsilon_F)=2\pi\hbar \int \frac{d^2k}{(2\pi)^2}\delta(\epsilon_F-\epsilon_k)\Omega_z(k)\cos 3\theta_k$, where $\hbar$ is the reduced Planck constant, $\epsilon_k$ is the energy dispersion, $\epsilon_F$ is the Fermi energy and $\theta_k$ is the angle measured from the $\Gamma$-K line[8]. According to skew scattering theory (Eq. 2), the second-order transverse conductivity $\sigma^{(2)}_{yxx}$ is proportional to the inverse of the skew scattering time $\tilde{\tau}^{-1}$, which is proportional to $T(\epsilon_F)$ in the system with threefold symmetry[7]. Consequently, $\sigma^{(2)}_{yxx}$ is proportional to $T(\epsilon_F)$. In Fig. 4d, we further calculated the dependence of $T(\epsilon_F)$ on the Fermi energy ranging from 0 to 0.3 eV. Notably, when the band filling approaches the band crossing in Ta $L_{z,-}$ orbital near the M point, the sign of $T(\epsilon_F)$ changes from positive to negative at $\epsilon_F$ = 0.24 eV, indicating the Berry curvature triple strongly depends on the position of the Fermi level relative to key features in the band structure. Based on the experimental data shown in Figs. 2c and 2d, we further examine the dependence of $\sigma^{(2)}_{yxx}$ on $n_{total}$ under irradiation at 10 K, as presented in Fig. 4e. Red markers represent the regime dominated by a single carrier type, while blue markers correspond to the presence of two carrier types. The $\sigma^{(2)}_{yxx}$-$n_{total}$ relationship clearly shows that $\sigma^{(2)}_{yxx}$ undergoes a sign reversal from positive to negative as $n_{total}$ increases, i.e., as the band filling increases. Combining theoretical calculations with experimental results, we attribute the light-induced sign reversal of $\sigma$

$\sigma^{(2)}_{yxx}$ to the sign change in the Berry curvature triple, which originates from the Fermi-level-dependent Berry curvature distribution and can be effectively tuned via photoexcitation.

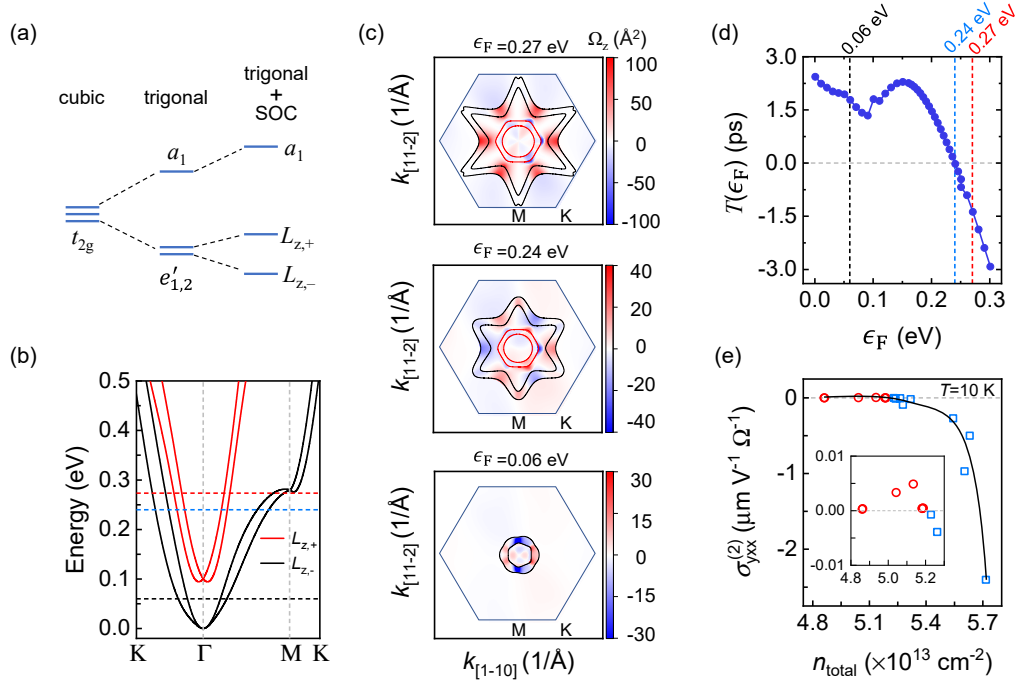

**Fig. 4 | Density functional theory calculations of the band structure and Berry curvature distribution.** (a) Energy-level diagram of Ta 5$d$ orbitals under a trigonal crystal field with $C_{3v}$ symmetry of the KTO (111) surface. The threefold degenerate $t_{2g}$ orbitals are split into a non-degenerate $a_1$ singlet and a twofold degenerate $e'$ doublet, and $e'$ states are further split into $L_{z,-}$ and $L_{z,+}$ orbitals due to SOC. (b) Calculated electronic band structure, including the lower $L_{z,-}$ (black) and upper $L_{z,+}$ (red) orbitals, for 6-Ta layers of the KTO (111) surface, obtained from first-principles calculations with a Hubbard $U_{eff}$ = 4.0 eV. (c) Fermi surfaces and corresponding Berry curvature distributions in the $k$ space at $\epsilon_F$ = 0.06, 0.24, and 0.27 eV, which are indicated by the dashed black, blue and red dashed lines in (b). The color bar shows the magnitude of the Berry curvature. (d) Calculated Berry curvature triple $T(\epsilon_F)$ as a function of $\epsilon_F$. (e) Dependence of $\sigma^{(2)}_{yxx}$ on $n_{total}$ under irradiation at $T$=10 K. The inset shows a magnified view of the $\sigma^{(2)}_{yxx}$ near zero. The red markers represent the regime with a single carrier type, while the blue markers indicate the presence of two types of carriers. Solid lines in (d) and (e) are guides for the eye.

In summary, we have demonstrated a giant and highly tunable nonlinear Hall effect in the 2DEG at the CZO/KTO (111) heterointerface under illumination. The photoinduced second-order transverse conductivity $\sigma^{(2)}_{yxx}$ not only undergoes a sign reversal but also increases by nearly five orders of magnitude, highlighting the strong optical control over nonlinear Hall response. Scaling analysis indicates that the NLHE is dominated by the skew scattering mechanism, exhibiting a positive contribution in the low-$\sigma_{xx}^2$ regime with a single carrier type, and undergoing a sign reversal with a significantly larger negative contribution in the high-$\sigma_{xx}^2$ regime due to photoexcited high-mobility carriers. According to skew scattering theory, the giant enhancement of $\sigma_{yxx}^{(2)}$ arises from an increased cubic transport scattering time ($\tau^3$)—associated with high-mobility photocarriers. DFT calculations further indicate that photoexcitation shifts the Fermi level upward, populating the higher-lying Ta $L_{z,+}$ orbital and alerting the Berry curvature distribution. Notably, the sigh change of the BCT as the Fermi level approaches a band crossing in the Ta $L_{z,-}$ orbital near the M point further explains the observed sign reversal of $\sigma_{yxx}^{(2)}$, underscoring the high sensitivity of nonlinear transport to the band structure. These findings reveal a new route for light-controlled nonlinear electronic responses in oxide heterostructures and provide insights into Berry curvature engineering via photoexcitation in symmetry-broken quantum systems.

## METHODS

### Sample and device fabrication

CZO film with the thickness of 10 nm was grown on 5×5×0.5 mm$^3$ (111)-oriented KTO single crystalline substrate by the technique of pulsed laser deposition (PLD) with a ceramic CZO target. A KrF Excimer laser (wavelength is 248 nm) was employed. The repetition rate was 2 Hz and the fluence was ~2 J/cm$^2$. During the deposition process, the substrates temperature was kept at 750 °C, and the oxygen partial pressure was fixed at 5×10$^{-3}$ Pa. After deposition, the sample was furnace-cooled to room temperature under the same atmosphere. Film thickness was determined by the number of laser pulses, which has been carefully calibrated by small-angle X-ray reflectivity. Surface morphology of the heterostructure was measured by atomic force microscope (AFM, MultiMode 8, Bruker). For electrical measurements, the Hall-bar devices were fabricated using the standard photolithography followed by argon ion beam etching. The central Hall-bar channel has a width of 10 μm and a length of 60 μm.

### Electrical transport measurements

The standard six-probe configuration (Fig. 1b) was used for the transport characterization through a commercial physical property measurement system (PPMS, Quantum Design). Ultrasonic wire bonding with 20 μm diameter Al wires was employed for electrical connections. We performed low-frequency AC harmonic electric measurements, using Keithley 6221 current source and Stanford Research SR830 lock-in amplifiers. During the measurements, a sinusoidal current with a constant amplitude and certain frequency was applied to the devices, and the in-phase (0°) first harmonic $V^{\omega}$ and out-of-phase (90°) second harmonic $V^{2\omega}$ longitudinal and transverse voltages were measured simultaneously by two lock-in amplifiers. To investigate the effect of photoexcitation on transport measurements, semiconductor laser beams ($\lambda$= 375, 405, 532, and 660 nm) were introduced into PPMS by an optical fiber to illuminate the devices. The laser spot size of the light on devices was ≈ 2 mm in diameter. The laser spot on the device surface had a diameter of approximately 2 mm. The laser power (0~15 mW) and the corresponding photon flux (0~9.01×10$^{17}$ cm$^{-2}$ s$^{-1}$) were calibrated at the output of the optical fiber. All light-controlled measurements were performed after a sufficient waiting period to allow the resistance to stabilize under illumination. Furthermore, after each cycle of light irradiation at a given wavelength, the CZO/KTO heterostructure was warmed to room temperature to restore their initial state, thereby ruling out persistent photoconductivity.

### First-principles calculation

We performed the DFT first-principles calculations using a plane wave basis set and projector augmented wave method[36], as implemented in the Vienna *ab initio* Simulation Package (VASP)[37,38]. A slab model of 6 units cell KTO on the (111) surface and an in-plane experimental lattice constant of $\sqrt{2}a_{KTO}$, a Γ-centered (6 × 6 × 1) k-point mesh and an energy cutoff of 500 eV have been used. The standard Perdew-Burke-Ernzerhof pseudopotential[39] was adopted in all calculations, treating 9 valence electrons for K ($3s^23p^64s^1$), 11 valence electrons for Ta ($5p^65d^46s^1$) and 6 valence electrons for O ($2s^22p^4$). To account for the correlation effects of the Ta 5$d$ electrons, the rotationally invariant Dudarev's formalism was performed with the reported $U_{\text{eff}}$ = 4.0 eV[16,40,41]. Atomic positions were optimized until the Hellmann-Feynman force on each atom was smaller than 0.01 eV Å$^{-1}$ and the electronic iteration was performed until the total energy change was smaller than 10$^{-7}$ eV. The SOC was included in self-consistent calculations and band calculations. To calculate the Berry curvature, we constructed the Wannier tight-binding Hamiltonian obtained from the Wannier90 code[42-44]. In the construction, 336 Wannier functions, including the Ta-$d$ and O-$p$ orbitals, were chosen. The Wannier interpolation approach with 601 × 601 crystal momentum points was adopted in WannierTools[45] for calculations of the Berry curvature and the Berry curvature triple.

**Data availability**
The authors declare that data generated in this study are provided in the paper and the Supplementary Information file. Further datasets are available from the corresponding author upon request.

**Acknowledgments**
This work has been supported by the Science Center of the National Science Foundation of China (Grant No. 52088101), the National Key Research and Development Program of China (Grant No. 2022YFA1403302, No. 2024YFA1410200, No. 2021YFA1400300, No. 2021YFB3501200, No. 2021YFB3501202, and No. 2023YFA1406003), the National Natural Science Foundation of China (Grant No. 12474103, No. T2394470, No. T2394474, No. 12274443, No. 92263202, No. U23A20550, and No. 22361132534), the Strategic Priority Research Program B of the Chinese Academy of Sciences (Grant No. XDB33030200), and the Beijing Outstanding Young Scientist Program. The authors acknowledge the Center for Micro-Nano Innovation of Beihang University for the sample morphological characterization.

**Author contributions**
H.Z. conceived the project and proposed the strategy. H.Z., J.R.S., and W.S.Z. supervised the study. D.M.T. and L.C. grew the samples and fabricated the devices. D.M.T., W.J.Q., M.L., and Y.T.B. developed the experimental setup and performed the transport measurements. X.B.C. carried out the theoretical calculations. H.Z., D.M.T., X.B.C., and J.R.S. analyzed the experimental and theoretical data. J.E.Z., F.R.H., H.W.Y., Y.S.C., Y.Z.C., F.X.H., J.W., Y.B.X., and B.G.S. contributed to data analysis and discussion. H.Z. and D.M.T. wrote the manuscript with input from all authors. All authors contributed to manuscript revision.

**Conflict of Interest**
The authors declare no conflict of interest.